\documentclass[a4paper,12pt]{myarticle}
\usepackage{graphicx}
\renewcommand\figurename{\small {\sc Fig.}}
\renewcommand\tablename{\small {\sc Table}}
\begin{document}
\thispagestyle{empty}
\ \vspace{0cm}
\begin{center}
\renewcommand{\baselinestretch}{1.1} \normalsize
{\bf \large Estimated Correlation Matrices and Portfolio Optimization} \par
\vspace{0.7cm}
{Szil\'ard Pafka$^{1,2}$ and Imre Kondor$^{1,3}$} \par \vspace{0.35cm}
{\it \small $^1$Department of Physics of Complex Systems, E\"otv\"os
University \\
P\'azm\'any P.\ s\'et\'any 1/a, H--1117 Budapest, Hungary} \par
\vspace{0.1cm}
{\it \small $^2$Risk Management Department, CIB Bank \\
Medve u.\ 4--14., H--1027 Budapest, Hungary} \par
\vspace{0.1cm}
{\it \small $^3$Institute for Advanced Studies, Collegium Budapest  \\
Szenth\'aroms\'ag u. 2., H--1014 Budapest, Hungary} \par \vspace{0.45cm}
April, 2003 \par \vspace{1cm}
{\bf Abstract} \par \vspace{0.7cm}
\parbox{13cm}{\small
Financial correlations play a central role in financial 
theory and also in many practical applications. From theoretical point of view,
the key interest is in a proper description of the structure and 
dynamics of correlations. From practical point of view, the
emphasis is on the ability of the developed models to 
provide the adequate input for the numerous portfolio and risk
management procedures used in the financial industry. This is crucial,
since it has been long argued that correlation matrices determined from
financial series contain a relatively large amount of noise and,
in addition, most of the
portfolio and risk management techniques used in practice
can be quite sensitive to the inputs.
In this paper we introduce a model (simulation)-based approach 
which can be used for a systematic investigation of the effect of the
different sources of noise in financial correlations in the portfolio 
and risk management context.
To illustrate the usefulness of this framework, we develop several
toy models for the structure of correlations and, by considering the
finiteness of the time series as the only source of noise,
we compare the performance of several correlation matrix estimators
introduced in the academic literature and which have since gained also
a wide practical use. 
Based on this experience, we believe that our simulation-based approach
can also be useful for the systematic investigation of several other problems
of much interest in finance.
\par \vspace{0.3cm} {\it PACS:} 87.23.Ge; 05.45.Tp; 05.40.--a
\par \vspace{0.3cm} {\it Keywords:} financial correlation matrices,
estimation error, random matrix theory, noise filtering, 
portfolio optimization, capital allocation, risk management
} \end{center} \vspace{1cm} \par
\rule{5cm}{0.4pt} \par
{\small {\it E-mail:} syl@complex.elte.hu (S.\ Pafka),
kondor@colbud.hu (I.\ Kondor)
\newpage \setcounter{page}{1}
\renewcommand{\baselinestretch}{1.1} \normalsize


\section{Introduction}

Correlation matrices of financial returns play a crucial role in
several branches of modern finance such as investment theory, 
capital allocation and risk management. For example, financial
correlation matrices are the key input parameters to Markowitz's
classical portfolio optimization problem \cite{markowitz}, which
aims at providing a recipe for the selection of a portfolio of
assets such that risk (quantified by the standard deviation of
the portfolio's return) is minimized for a given level of expected
return. For any practical use of the theory it would therefore be
necessary to have reliable estimates for the correlations of returns
(of the assets making up the portfolio), which are usually obtained
from historical return series data. However, if one estimates
a $n\times n$ correlation matrix from $n$ time series of length $T$ each,
since $T$ is usually bound by practical reasons, one inevitably
introduces estimation error, which for large $n$ can become so overwhelming
that the whole applicability of the theory may become questionable.

This difficulty has been well known by economists for a long time
(see e.g.\ \cite{eltongruber} and the numerous references therein).
Several aspects of the effect of noise (in the correlation matrices 
determined from empirical data) on the classical portfolio selection 
problem has been investigated e.g.\ in refs.\ \cite{econnoisedet}.
One way to cope with the problem of noise is to impose some structure on
the correlation matrix, which may certainly introduce some bias in the 
estimation, but by reducing effectively the dimensionality of the problem,
could be in fact expected to improve the overall performance of the estimation. 
The best-known such structure is that imposed by the single-index (or market) model,
which has gained a large interest in the academic literature (see e.g.\ 
\cite{eltongruber} for an overview and references) and has also
become widely used in the financial industry (the coefficient "beta", relating
the returns of an asset to the returns of the corresponding wide market index,
has long become common talk in the financial
community). On economic or statistical grounds, several other 
correlation structures have been experimented with in the academic
literature and financial industry, for example multi-index
models, grouping by industry sectors, macroeconomic factor models, models 
based on principal component analysis etc. Several studies (see e.g.\ refs.\
\cite{eltongruberpap}) attempt to compare the performance of these
correlation estimation procedures as input providers for the 
portfolio selection problem, although all these studies have been somewhat 
restricted to
the use of given specific empirical samples. More recently, other procedures
to impose some structure on correlations (e.g.\ Bayesian shrinkage estimators)
or bounds directly on the portfolio weights (e.g.\ no short selling) 
has been explored, see e.g.\ refs.\ \cite{others}. The general conclusion 
of all these studies is that reducing the dimensionality of the problem
by imposing some structure on the correlation matrix may be of great help for
the selection of portfolios with better risk--return characteristics.

The problem of estimation noise in financial correlation matrices has
been put into a new light by \cite{galluccio,bouchaud,stanley} from
the point of view of random matrix theory. These studies have shown that
empirical correlation matrices deduced from financial return series
contain such a high amount of noise that, apart from a few
large eigenvalues and the corresponding eigenvectors, their structure
can essentially be regarded as random. In \cite{bouchaud}, e.g., it is
reported that about 94\% of the spectrum of correlation matrices determined
from return series of the S\&P 500 stocks can be fitted by that of a
random matrix. Furthermore, two subsequent studies 
\cite{bouchaud2,stanley2} have
shown that the risk--return characteristics of optimized portfolios could 
be improved if prior to optimization one filtered out the lower part of the 
eigenvalue spectrum of the correlation matrix in an attempt to remove 
(at least partially) the noise, a procedure similar to principal 
component analysis. 
Other approaches inspired from physics and that are aimed to be useful in 
extracting information from
noisy correlation data have been introduced in \cite{mantegna,kertesz}.
It is important to note that all the above studies have 
used (given) empirical datasets,
which in addition to the noise due to the finite length of the time
series, contain also several other sources of error (caused by 
non-stationarity, market microstructure etc.).

The motivation of our previous study \cite{noisy2} came from this 
context. In order to get rid of these additional sources of errors,
we based our analysis on data artificially generated from some
toy models. This procedure offers a major advantage in that the "true"
parameters of the underlying stochastic process, hence also the
correlation matrix is exactly known. The key observation of \cite{noisy2}
is that the effect of noise strongly depends on the ratio $T/n$,
where $n$ is the size of the portfolio while $T$ is the length of the
available time series. Moreover, in the limit $n\to\infty$,
$T\to\infty$ but $T/n=\textrm{const.}$ the suboptimality of the
portfolio optimized using the "noisy" correlation matrix (with respect
to the portfolio obtained using the "true" matrix) is $1/\sqrt{1-n/T}$
exactly.
Therefore, since the length of the time series $T$ is limited in any practical
application, any bound one would like to impose on the effect of noise
translates, in fact, into a constraint on the portfolio size $n$.

The aim of this paper is (besides to extend the analysis of the previous study)
to introduce a {\it model (simulation)-based approach} that can be generally
used for the systematic 
investigation of correlations in financial markets and for the study of the effect 
of different sources of noise on the numerous procedures based on correlation 
matrices extracted from financial data. 
As an illustration of the usefulness of this approach, we introduce several toy
models aimed to progressively incorporate the relevant features of 
real-life financial correlations and, in the world of these models, we study the
effect of noise (in this case only due to the estimation error caused by the 
finiteness of surrogate time series generated by the models) on the classical
portfolio optimization problem. More precisely, we compare the performance of 
different
correlation matrix "estimation" methods (e.g.\ the filtering procedure introduced
in \cite{bouchaud2,stanley2}) in providing inputs for the selection of 
portfolios with optimal risk--return characteristics. 
The approach is in fact very common in physics, where one starts with some
bare model and progressively adds finer and finer elements, while studying
the behavior of the "world" embodied by the model by comparing it to the
real-life (experimental) results.
We strongly believe that our
model-based approach can be useful for the 
{\it systematic} study of several other 
problems in which financial correlation matrices play a crucial role.


\section{Results and Discussion}

We keep to consider the following simplified version of the classical 
portfolio optimization problem introduced in \cite{noisy2}: 
the portfolio variance $\sum_{i,j=1}^n w_i\,\sigma_{ij}\,w_j$
is minimized under the budget constraint $\sum_{i=1}^n w_i=1$, where
$w_i$ denotes the weight of asset $i$ in the portfolio and
$\sigma_{ij}$ the covariance matrix of returns. 
This simplified form provides the most convenient laboratory 
for testing the effect of noise in correlations,
since it eliminates the additional uncertainty arising from the determination of
several other parameters that appear in more complex formulations. The weights 
of the optimal portfolio in this simple case are:
\begin{equation}
\label{eq:sol}
w_i^*=\frac{\sum_{j=1}^n \sigma_{ij}^{-1}}{\sum_{j,k=1}^n \sigma_{jk}
^{-1}}.
\end{equation}

Starting from a given "true" covariance matrix $\sigma_{ij}^{(0)}$ ($n\times n$)
we generate surrogate time series $y_{it}$ (of finite length $T$),
$y_{it}=\sum_{j=1}^n L_{ij}\,x_{jt}$, with
$x_{jt}\sim \textrm{i.i.d.\ N}(0,1)$ and $L_{ij}$ the
Cholesky decomposition of the matrix $\sigma_{ij}^{(0)}$.
In this way we obtain "return series" $y_{it}$ that have a
distribution characterized by the "true" covariance matrix $\sigma_{ij}^{(0)}$.
Similar to real-life situations (where the true covariance matrix is not known) 
we calculate different "estimates" $\sigma_{ij}^{(1)}$
for the covariance matrix based on several 
competing procedures and then use these estimates in our portfolio optimization.
Finally, we compare the performance of these procedures using metrics related to
the risk (standard deviation) of the "optimal" portfolios constructed based on
the corresponding estimates.
The main advantage of this simulation-based approach is that
the "true" covariance matrix can be incorporated in the evaluation, which is
certainly much cleaner than using, as in empirical studies, some proxy for it 
(which in turn introduces an additional source of noise).

In our previous study \cite{noisy2} we have used a very simple structure ("model") for 
$\sigma_{ij}^{(0)}$ (namely the identity matrix) and we have studied the effect of noise 
when the "estimated" matrix $\sigma_{ij}^{(1)}$ is the sample (or historical)
covariance matrix. In this paper we introduce several other "models" 
(proposals for the structure of $\sigma_{ij}^{(0)}$) 
which are intended to incorporate progressively
the most relevant characteristics of real-life financial correlations
(the models are given in terms of the corresponding 
correlation matrix $\rho_{ij}^{(0)}$):

\begin{enumerate}
\item "Single-index", "market" or "average correlation" model. The
correlation matrix has 1 in the diagonal
and $\rho_0$ given ($0<\rho_0<1$) off-diagonal (all correlations the same, 
hence the name of "average correlation" model). 
The eigenstructure of such a matrix is formed of one 
large\footnote{$\lambda_1=1+(n-1)\rho_0$, which for the usual values
of the parameters is large compared to
$\lambda_2=\lambda_3=\ldots=\lambda_n=1-\rho_0$.} 
eigenvalue with corresponding eigenvector in the direction
of $(1,1,\ldots,1)$ and a $n-1$-fold degenerated eigenvalue subspace orthogonal
on the subspace of the first eigenvector. The eigenspace of the large eigenvalue
can be thought of as describing correlations with a broad "index" composed of all
stocks (the "market"), hence the name of "single-index" or "market" model.
This model is motivated by the similar salient feature of stock market correlations
found by numerous research studies (see e.g.\ \cite{eltongruber} for references).

\item "Market+sectors" model. A very simple structure intended to incorporate
this more debated\footnote{See e.g.\ refs.\ \cite{industry}.} feature of real-life 
financial correlations can be based
on a correlation matrix composed of $n_1\times n_1$ blocks (with 1 in the 
diagonal and $\rho_1$ off-diagonal) and $\rho_0$ outside the blocks 
($0<\rho_0<\rho_1<1$ and $\frac{n}{n_1}$ integer). In this model there is still
a strong influence of the "market" but stocks from the same
block ("industrial sector") display additional common 
correlations. On the other hand, the eigenspectrum of such a 
matrix\footnote{The eigenstructure is formed of 
a large eigenvalue $\lambda_1=1+(n_1-1)\rho_1+(n-n_1)\rho_0$, a $\frac{n}{n_1}-1$-fold
degenerated subspace corresponding to medium-size eigenvalues 
$\lambda_2=\lambda_3=\ldots=\lambda_{\frac{n}{n_1}}=1+(n_1-1)\rho_1-n_1\rho_0$ and a
$n-\frac{n}{n_1}$-fold degenerated subspace with eigenvalues
$\lambda_{\frac{n}{n_1}+1}=\lambda_{\frac{n}{n_1}+2}=\ldots=\lambda_n=1-\rho_1$.}
is closer to the eigenspectrum of real-life financial correlation matrices 
as described e.g.\ in \cite{stanley2}. This correlation structure also fits
better with the findings of \cite{mantegna,kertesz}, which using a
hierarchial tree approach found also that stocks tend to be coupled according
to their belonging to industrial sectors.

\item "Semi-empirical" (bootstrapped) model. Starting from a large set of 
empirical financial data\footnote{The same dataset as in 
\cite{noisy2} has been used. We thank again J.-P.\ Bouchaud and L.\ Laloux for 
making their data \cite{bouchaud,bouchaud2} available to us.}
for each portfolio size $n$, we select randomly (bootstrap) $n$ time series 
from the set of empirical return data and
an $n\times n$ covariance matrix is calculated using the full length
of the available series. This matrix is then used as $\sigma_{ij}^{(0)}$
in the simulations (to generate the surrogate data). In order to examine
the sensitivity of our results with respect to the choice of the $n$ time
series, we repeat several times the simulations (with different bootstrapped
empirical series) and we compare the results.
The correlation structure of this model is hoped to be the closest to
real-world financial correlations, although the disadvantage of it is,
similar to empirical studies, that it is based on a given set of empirical
data which might be representative in certain situations but it is still
not fully general.
\end{enumerate}

In the framework of each of the models introduced above, 
we investigate the performance of three alternative choices
for the "estimated" covariance matrix $\sigma_{ij}^{(1)}$:
\begin{enumerate}
\item Sample (historical) covariance matrix. 

\item "Single-index" covariance matrix, i.e.\ the matrix obtained from
the sample covariance matrix by a simplified filtering procedure similar to 
the one described below, but considering only the largest eigenvalue
(and the corresponding eigenvector), which is believed to correspond to
a broad market index covering all stocks, see e.g.\ \cite{stanley2}.

\item Filtered covariance matrix using the procedure based on
random matrix theory \cite{bouchaud2,stanley2}. For this, one starts
with the sample correlation matrix and keeps only the eigenvalues and the
corresponding eigenvectors reflecting deviations from random matrix theory
predictions (those outside the random matrix noise-band) and then constructs 
a "cleaned" correlation matrix such that the trace of the matrix is preserved.
The intuition behind this procedure is that deviations from random matrix
theory predictions should correspond to "information" and describe genuine
correlations in the system while the eigenstates corresponding to random matrix 
theory predictions should be manifestations of purely random "noise". 
The filtered covariance matrix is then obtained from the filtered
correlation matrix and sample standard deviations. This procedure 
is very much reminiscent of principal component analysis, although classical
multivariate analysis gives generally no hints about how many components
(factors) to include in the matrix constructed using the principal components
(see e.g.\ \cite{multivanal}).
The filtering procedure based on random matrix theory can therefore be thought of as
a theoretically sound indication for the number of principal components
to be included in the analysis.
\end{enumerate}

To study the effect of noise on the portfolio optimization
problem we use metrics based on the following quantities:
\begin{enumerate}
\item $\sum_{i,j=1}^n w_i^{(0)*}\,\sigma_{ij}^{(0)}\,w_j^{(0)*}$,
the "true" risk of the optimal portfolio without noise, where
$w_i^{(0)*}$ denotes the solution to the optimization problem
with $\sigma_{ij}^{(0)}$;
\item $\sum_{i,j=1}^n w_i^{(1)*}\,\sigma_{ij}^{(0)}\,w_j^{(1)*}$,
the "true" risk of the optimal portfolio determined in the
case of noise, where $w_i^{(1)*}$ denotes the solution to the
optimization problem with $\sigma_{ij}^{(1)}$;

\item $\sum_{i,j=1}^n w_i^{(1)*}\,\sigma_{ij}^{(1)}\,w_j^{(1)*}$,
the "predicted" risk (cf.\ \cite{bouchaud2,stanley2,noisy2}), that is
the risk that {\it can be} observed when the optimization is based on
the "empirical" series;

\item $\sum_{i,j=1}^n w_i^{(1)*}\,\sigma_{ij}^{(2)}\,w_j^{(1)*}$,
the "realized" risk (cf.\ \cite{bouchaud2,stanley2,noisy2}), that is
the risk that {\it would be} observed if the portfolio were held one more
"period", where $\sigma_{ij}^{(2)}$ is the covariance
matrix calculated from the returns in this second period.
\end{enumerate}
To facilitate the comparison, we calculate the ratios of the square
roots of the three latter quantities to the first one, and
denote these by $q_0,q_1$ and $q_2$, respectively.
That is $q_0,q_1$ and $q_2$ represent the "true", the "predicted" resp.\
the "realized" risk, expressed in units of the "true" risk in the
absence of noise. In other words, $q_0$ describes directly the ability of
a given estimation procedure to provide the correct input for portfolio
optimization, $q_1$ describes the bias one makes if then uses the estimated
matrix for the calculation of the risk of the optimal portfolio, while 
$q_2$ is the risk measured if one waits in time and uses
the information from the new series for risk measurement (see also
\cite{noisy2}).

We start with presenting the simulation results when the series have been
generated using the {\sl "market" model} (for $\sigma_{ij}^{(0)}$). Since the
main feature of the correlation structure (one outstanding large eigenvalue) is, 
at least for
the parameter values used in our simulations, preserved also in the correlation
matrix obtained from the generated series ($\sigma_{ij}^{(1)}$), the results
for the filtering based on the largest eigenvalue and on random matrix
theory are in fact the same. Therefore, we proceed with comparing the performance 
of the historical and filtered estimation procedures for different values of
the model parameters $n$, $T$ and $\rho_0$ using the evaluation metrics 
$q_0$, $q_1$, $q_2$ and $q_2/q_1$. 
A summary of our simulation results is presented in Table \ref{tbl:simulsimple}.

\begin{table}[h!]
\caption{Optimal portfolio risk and performance indicators
for the historical ($h$) and market ($m$) correlation matrix estimators for different
values of the parameters of the model ($\sigma_{ij}^{(0)}$).
\label{tbl:simulsimple}} 
\begin{tabular}{|c|c|c|c||c|c|c|c|c|c|c|c|} \hline 
$\rho_0$ & $n$ & $T$ & $T/n$ & $q_0^{(h)}$ & $q_0^{(m)}$ & $q_1^{(h)}$ & $q_1^{(m)}$ &
 $q_2^{(h)}$ & $q_2^{(m)}$ & {\footnotesize $q_2/q_1^{(h)}$} & 
{\footnotesize $q_2/q_1^{(m)}$} \\ \hline
0.2 & 200 & 300 & 1.5 & 1.77 & 1.11 & 0.56 & 0.78 & 1.77 & 1.13 & 3.16 & 1.46 \\ \hline
0.2 & 1000 & 1500 & 1.5 & 1.73 & 1.12 & 0.59 & 0.78 & 1.71 & 1.11 & 2.96 & 1.42 \\ \hline
0.6 & 1000 & 1500 & 1.5 & 1.75 & 1.11 & 0.58 & 0.77 & 1.75 & 1.12 & 3.01 & 1.45 \\ \hline \hline
0.2 & 1000 & 2000 & 2 & 1.42 & 1.11 & 0.71 & 0.82 & 1.43 & 1.11 & 2.00 & 1.35 \\ \hline
0.2 & 1000 & 5000 & 5 & 1.11 & 1.07 & 0.89 & 0.91 & 1.12 & 1.07 & 1.26 & 1.18 \\ \hline \hline
0.2 & 1000 & 500 & 0.5 & - & 1.12 & - & 0.57 & - & 1.12 & - & 1.92 \\ \hline
\end{tabular}
\end{table}

It turns out that, for sufficiently large $n$ and $T$, the value of the $q$'s
depends strongly only on $T/n$ (and, interestingly, does not seem to depend  
on $\rho_0$). This can be seen also from the results presented in Table 
\ref{tbl:simulsimple} (the variation in the first 3 rows is in fact within
the usual standard deviation bounds).
This is not very surprising as concerning the results for the
historical matrix, which has been studied in our previous paper \cite{noisy2}.
The strong dependence on $T/n$ seems to be valid, however, also when the
filtered matrix is used. One important difference to note is, however, the
significant improvement in the risk characteristics of the optimal portfolio
when the filtering procedure is used for estimation, e.g.\ for $T/n=2$ instead
of obtaining a portfolio with risk more than 40\% larger than the trully
optimal one (see $q_0$), using the filtering procedure one can get portfolios with risk
only 10\% larger. Furthermore, as it can also be seen from the table,
using the filtered matrix one can obtain portfolios
close to the optimal one even for $T\leq n$ when the sample (historical) matrix is
singular and not at all appropriate for being used in the optimization. 
This improvement in performance is not difficult to understand, since
with the filtering procedure one implicitly incorporates into the "estimation" 
the additional information about the structure of the correlation matrix.
Note also that for all parameter values $q_2$ is very close to $q_0$, therefore
the risk measured in the second "period" seems to be a good proxy for the
"true" risk of the optimal portfolio.

\begin{table}
\caption{Optimal portfolio risk and performance indicators 
for the historical ($h$), market ($m$) and random matrix theory ($r$) 
correlation matrix estimators for different
values of the parameters of the model ($\sigma_{ij}^{(0)}$).
\label{tbl:simulparisi}} 
{\small
\begin{tabular}{|c|c|c|c|c||c|c|c|c|c|c|c|c|c|} \hline 
$\rho_0$ & $\rho_1$ & $n_1$ & $n$ & $T$ & $q_0^{(h)}$ & $q_0^{(m)}$ & $q_0^{(r)}$ 
& $q_1^{(h)}$ & $q_1^{(m)}$ & $q_1^{(r)}$ & {\footnotesize $q_2/q_1^{(h)}$} & 
{\footnotesize $q_2/q_1^{(m)}$} & {\footnotesize $q_2/q_1^{(r)}$}\\ \hline
0.2 & 0.4 & 25 & 200 & 300 & 1.71 & 1.27 & 1.13 & 0.58 & 0.77 & 0.76 & 2.93 & 1.65 & 1.47 \\ \hline
0.2 & 0.4 & 25 & 1000 & 1500 & 1.75 & 1.28 & 1.13 & 0.58 & 0.77 & 0.76 & 3.07 & 1.63 & 1.46 \\ \hline
0.2 & 0.6 & 25 & 1000 & 1500 & 1.74 & 1.64 & 1.13 & 0.59 & 0.78 & 0.76 & 2.94 & 2.09 & 1.47 \\ \hline
0.4 & 0.6 & 25 & 1000 & 1500 & 1.73 & 1.36 & 1.13 & 0.58 & 0.77 & 0.76 & 2.96 & 1.77 & 1.49 \\ \hline
0.2 & 0.4 & 50 & 1000 & 1500 & 1.71 & 1.42 & 1.12 & 0.58 & 0.77 & 0.77 & 2.96 & 1.84 & 1.46 \\ \hline \hline
0.2 & 0.4 & 25 & 1000 & 2000 & 1.42 & 1.24 & 1.12 & 0.70 & 0.82 & 0.81 & 1.99 & 1.50 & 1.37 \\ \hline
0.2 & 0.4 & 25 & 1000 & 5000 & 1.11 & 1.16 & 1.07 & 0.89 & 0.91 & 0.90 & 1.24 & 1.27 & 1.17 \\ \hline \hline
0.2 & 0.4 & 25 & 1000 & 500 & - & 1.24 & 1.19 & - & 0.58 & 0.55 & - & 2.14 & 2.17 \\ \hline
\end{tabular}}
\end{table}

\begin{figure}
\begin{center}
\includegraphics[scale=0.6,angle=-90]{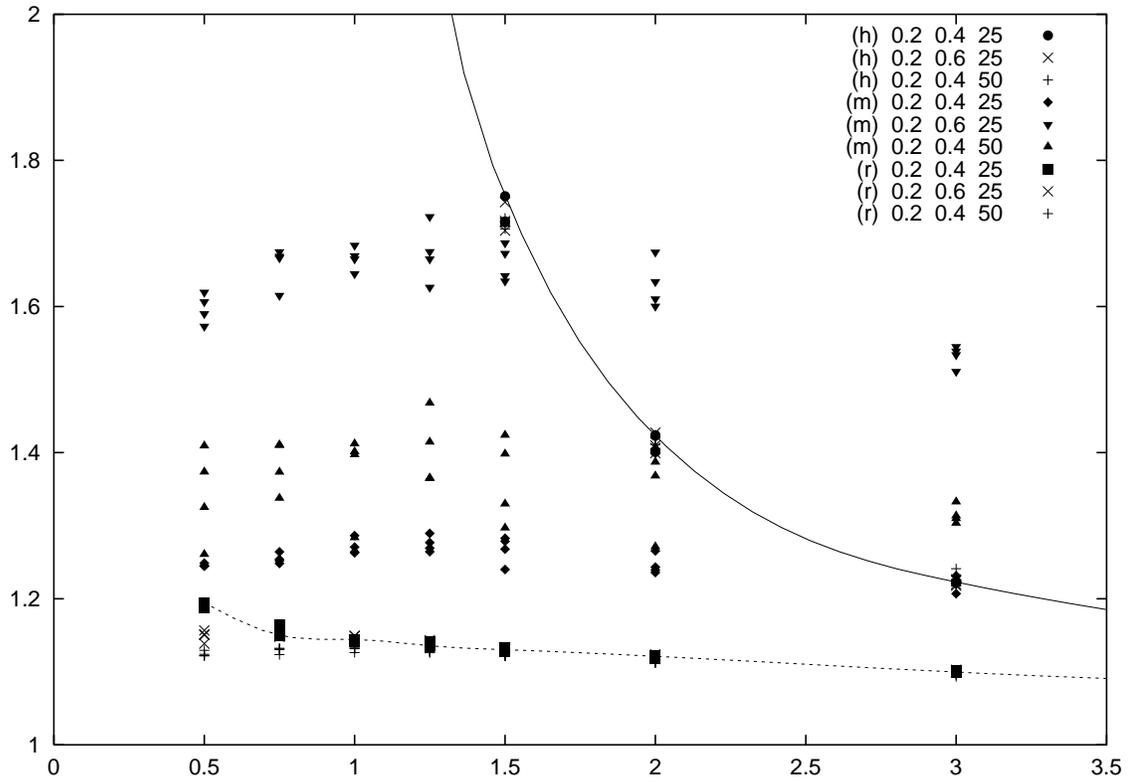}
\end{center}
\caption{$q_0$ as a function of $T/n$ for different values of the parameters
$\rho_0$, $\rho_1$ and $n_1$ and different values of $n$ and $T$. 
In the case of the historical and random matrix theory
estimator ($h$ and $r$, resp.) the points line up approximately on a line 
(solid and dotted, resp.). For the market estimator ($m$),
however, the dependence on virtually all the parameters is clear from the
figure (e.g.\ the increase in either $\rho_1$ or $n_1$ leads to the increase
of $q_0$).
\label{fig:simulparisi}}
\end{figure}

We next present the results when the series are generated with the
{\sl "market+sectors" model}, for different values of the parameters 
$n$, $T$, $n_1$, $\rho_0$ and $\rho_1$. Our results are summarized in
Table \ref{tbl:simulparisi}. The values for $q_2$'s have been again very
close to $q_0$ and therefore have been left out from the table.
We have found that the value of the $q$'s
in the case of the historical and random matrix theory-based estimators, again,
depends strongly on $T/n$ and not on the value of the other parameters,
while this is not true for the estimator based on the largest eigenvalue only.
This is illustrated in Fig.\ \ref{fig:simulparisi}, where $q_0$ in the
case of the three estimators is represented as a function of $T/n$ for
different value of the parameters $n$, $T$, $n_1$, $\rho_0$ and $\rho_1$.
The dependence of $q_0$ for the "single-index" estimator on the parameters
$\rho_0$, $\rho_1$ and $n_1$ can be easily understood, since either the increase
of $\rho_1$ or $n_1$, or the decrease of $\rho_0$ can be thought of as the
increase in the relative strength of "inter-sector" correlations (relative to
the overall correlation corresponding to the "market") and therefore an
estimator taking into account only the "market" component of correlations
(and ignoring the "sector" component) is of course expected to perform worse is
this case. 
Another important point to note is that, in most cases, the random matrix theory based
filtering outperforms the single-index estimator which in turn outperforms
the historical estimator. Moreover, the first two estimators can be used even
when the latter one provides a singular matrix totally inappropriate for input to the
portfolio optimization (for $T\leq n$).

Finally, we analyze the performance of the three correlation matrix estimators
in the case of the {\sl "semi-empirical" model} for $\sigma_{ij}^{(0)}$ (the matrix is
bootstrapped from the empirical matrix of a given large set 
of financial series). More precisely, for each value of the parameter $n$, we select 
at random $n$ series from the available dataset and we calculate the historical
matrix which is then used as $\sigma_{ij}^{(0)}$ in our 
simulations\footnote{Since most of the values for the length $T$ of the 
time series used in
our simulations is small compared to the lengths of the original dataset
from which $\sigma_{ij}^{(0)}$ is computed, the noise due to the "measurement error" 
of $\sigma_{ij}^{(0)}$ can be hoped to be small compared to the noise (deliberately)
introduced by the finiteness of $T$.}. 
Our results are summarized in Table \ref{tbl:simulempir}
(the values for $q_2$'s have been again left out of the table.) In this case, the
$q$'s for the two filtering matrix estimations do not depend so strongly on
$T/n$, some dependence on $n$ (and $T$) can also
be observed (see Fig.\ \ref{fig:simulempir}). It can be said again 
that, in general, the filtering procedures outperform 
significantly the historical matrix estimation, with the filtering based on the
random matrix theory approach performing the best.

\begin{table}[h!]
\caption{Optimal portfolio risk and performance indicators 
for the historical ($h$), market ($m$) and random matrix theory ($r$) 
correlation matrix estimators for different
values of the parameters of the model ($\sigma_{ij}^{(0)}$).
\label{tbl:simulempir}} 
\begin{tabular}{|c|c|c||c|c|c|c|c|c|c|c|c|} \hline 
$n$ & $T$ & $T/n$ & $q_0^{(h)}$ & $q_0^{(m)}$ & $q_0^{(r)}$ 
& $q_1^{(h)}$ & $q_1^{(m)}$ & $q_1^{(r)}$ & {\footnotesize $q_2/q_1^{(h)}$} & 
{\footnotesize $q_2/q_1^{(m)}$} & {\footnotesize $q_2/q_1^{(r)}$}\\ \hline
200 & 300 & 1.5 & 1.70 & 1.30 & 1.20 & 0.58 & 0.78 & 0.83 & 3.03 & 1.67 & 1.44 \\ \hline
300 & 450 & 1.5 & 1.74 & 1.48 & 1.24 & 0.58 & 0.76 & 0.84 & 2.99 & 1.94 & 1.45 \\ \hline \hline
300 & 600 & 2 & 1.41 & 1.50 & 1.21 & 0.71 & 0.77 & 0.90 & 2.02 & 1.95 & 1.35 \\ \hline
300 & 1500 & 2 & 1.12 & 1.53 & 1.15 & 0.89 & 0.80 & 0.96 & 1.26 & 1.92 & 1.21 \\ \hline \hline
300 & 150 & 0.5 & - & 1.41 & 1.33 & - & 0.76 & 0.73 & - & 2.02 & 1.85 \\ \hline
\end{tabular}
\end{table}

\begin{figure}[h!]
\begin{center}
\includegraphics[scale=0.6,angle=-90]{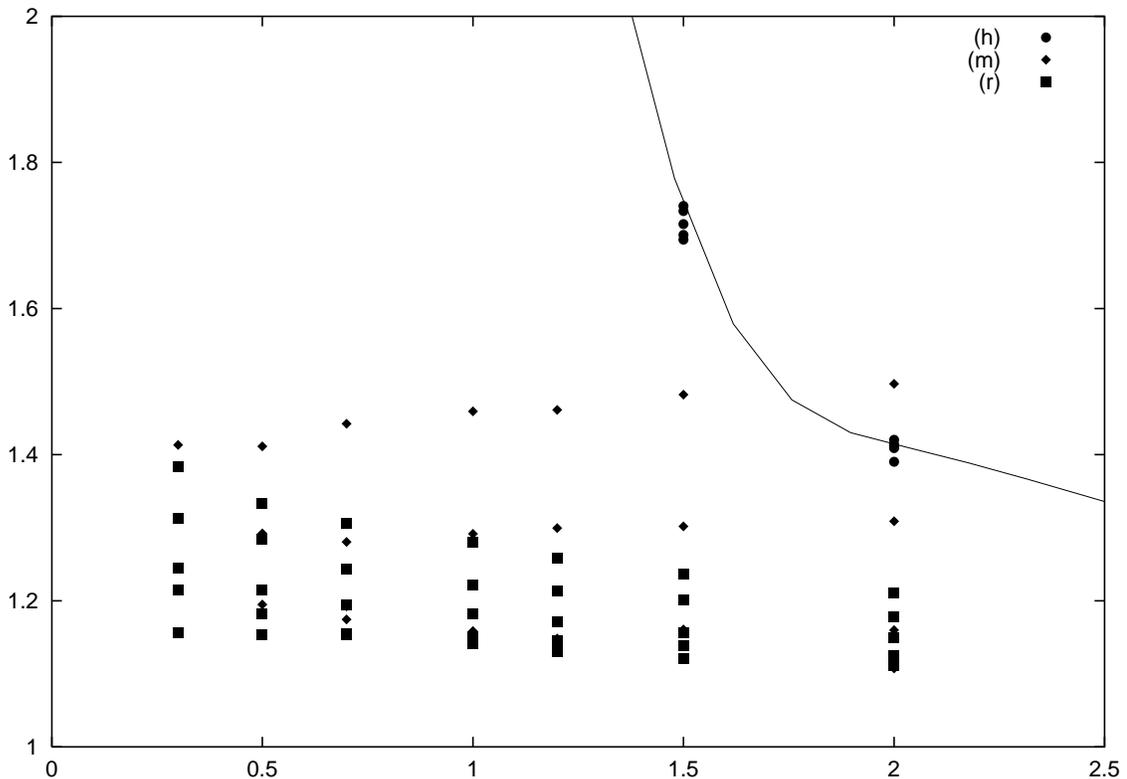}
\end{center}
\caption{$q_0$ as a function of $T/n$ for different values of $n$ and $T$. 
In the case of the historical estimator ($h$) the points line approximately 
on a line. For the market and random matrix theory estimator ($m$ and $r$,
resp.), however, the dependence on $n$ and $T$ is clear from the figure.
\label{fig:simulempir}}
\end{figure}

\newpage 
In conclusion, our simulation study provides a more general argument for
the usefulness of techniques for "massageing" empirical correlation matrices
before using them as inputs for portfolio optimization as suggested e.g.\ by
\cite{eltongruberpap,others,bouchaud2,stanley2}. Furthermore, it re-emphasizes
the fruitfulness of the random matrix theory-based filtering procedure for
portfolio selection applications.

There are several possibilities to extend the analysis of this paper.
One main direction would be to develop "models" that incorporate more
subtle features of real-life financial correlations. 
For example, an important feature of real financial series that has been
neglected is non-stationarity. 
Incorporating the dynamics of correlations into the model
could result into a more realistic description of correlations. For example,
models such as ARCH/GARCH and its numerous variants 
(see e.g.\ \cite{arch} for an overview) have been 
found to be fruitful in describing the dynamics of changing
volatility (and also of correlations in the multivariate setting). 
On the other hand, estimation 
techniques based on similar rationales (for example RiskMetrics
\cite{riskmetrics}) have been widely utilized by financial practitioners.
These estimation procedures 
run into the dimensionality problem typically already for $n=4$ or 5, 
but fortunately the principal component/factor approach has proved here also useful
\cite{factorarch}. 
A simple way to take account of non-stationarity in our "estimation"
would be to use exponential weighting of observations in the calculation
of the correlation matrix $\sigma_{ij}^{(1)}$ (in the spirit of RiskMetrics) and then
apply the filtering to this matrix. Of course, this should be preceded by 
the derivation of the corresponding formulae for the noise band of matrices with
this new structure.
Another way to extend the analysis of this study is to use the model 
(simulation)-based approach for evaluating the performance of several
other correlation matrix estimators introduced in the literature or
used in practice. 

The implications of successful noise filtering in correlation matrices used 
for portfolio optimization are enormous. Correlation matrices are not only
at the heart of modern finance and investment theory, but also appear in most
practical risk management and asset allocation procedures used in the financial
industry. In particular, most implementations of practical risk--return portfolio 
optimization or benchmark tracking (minimization of risk with respect to a 
given benchmark) involve either correlation matrices or "scenarios" usually generated 
using correlation matrices, see e.g.\ \cite{finind}. A short overview on the techniques
used by practitioners for reducing noise and estimation error in correlation 
matrices can be found in \cite{finindanal}. The filtering procedure based on
random matrix theory fits well into this package and can prove very useful for
reducing estimation error and its consequences.
On the other hand, from purely academic point of view, understanding the
structure and dynamics of correlations in financial markets is still of
central interest in finance and related fields, therefore any study that
makes it possible to reveal finer and finer bits of the structure of these 
correlations can be of great importance.


\section{Conclusion}

In this paper we introduced a model (simulation)-based approach
which can be used for a systematic investigation of the effect
of different sources of noise in correlation matrices determined
from financial return series. To show the usefulness of this approach
we developed several toy models for the structure of financial
correlations and, by considering only the noise arising from the
finite length of the model-generated time series, we analyzed the
performance of several correlation matrix estimation procedures 
in a simple portfolio optimization context. 

The results of this study can be extended in very numerous ways, some
of which are briefly given next.
First, by developing models that incorporate finer and finer elements
of the structure of financial correlations, the relevance of the results
can be increased further. For example, allowing for some dynamics
(non-stationarity) in correlations could make it possible to analyze
the effect of noise due to non-stationarity or due to the estimation
error of the parameters of some dynamic models on the portfolio optimization
problem.
Second, the analysis could be extended to several other correlation
estimation procedures introduced in the literature, e.g.\ trully
single-index model (with betas), multi-index models, different factor
estimation procedures, Bayesian-estimators etc.\ (see for example
\cite{eltongruberpap,others,finind,finindanal}).
The models (simulations) could also be
used for studying the performance of several other techniques for
the extraction of correlation information such as the hierarchial
tree methods of \cite{mantegna,kertesz}.
Third, our model-based approach can be used also in a more
complex optimization framework, e.g.\ in that of the classical mean--variance
efficient frontier rather than just in the simple global optimization framework
used in this paper.
Last, but not least, the approach could be used also for the study of 
different other more general
"correlation" measures if instead of the portfolio standard deviation
some other more sophisticated risk measure (e.g.\ Conditional Value-at-Risk)
is used.


\section*{Acknowledgements}

This work has been supported by the Hungarian National Science
Found OTKA, Grant No.\ T 034835.



\end{document}